\documentclass[singlecolumn,showpacs,preprintnumbers,amsmath,amssymb]{revtex4}

\usepackage{graphicx}
\usepackage{dcolumn}
\usepackage{bm}

\begin{document}


\title{Electrical conduction of silicon oxide containing silicon quantum dots}

\author{X. D. Pi}
\altaffiliation[Current address:\ \ ]{Department of Mechanical
Engineering, University of Minnesota, Minneapolis, Minnesota
55455}
\email{xdpi@umn.edu}
\author{O. H. Y. Zalloum}
\author{A. P. Knights}
\email{aknight@univmail.cis.mcmaster.ca}
\author{P. Mascher}
\affiliation{Department of Engineering
Physics, McMaster University, Hamilton, Ontario L8S 4L7, Canada}

\author{P. J. Simpson}
\affiliation{Department of Physics and Astronomy, University of
Western Ontario, London, Ontario N6A 3K7, Canada}

\date{\today}

\begin{abstract}
Current-voltage measurements have been made at room temperature on
a Si-rich silicon oxide film deposited via Electron-Cyclotron
Resonance Plasma Enhanced Chemical Vapor Deposition (ECR-PECVD)
and annealed at 750 - 1000$\ ^\circ$C. The thickness of oxide
between Si quantum dots embedded in the film increases with the
increase of annealing temperature. This leads to the decrease of
current density as the annealing temperature is increased.
Assuming the Fowler-Nordheim tunneling mechanism in large electric
fields, we obtain an effective barrier height $\phi_{eff}$ of
$\sim$ 0.7 $\pm$ 0.1 eV for an electron tunnelling through an
oxide layer between Si quantum dots. The Frenkel-Poole effect can
also be used to adequately explain the electrical conduction of
the film under the influence of large electric fields. We suggest
that at room temperature Si quantum dots can be regarded as traps
that capture and emit electrons by means of tunneling.
\end{abstract}

\maketitle

\newpage

\section{Introduction}
Si quantum dots (QDs) have been showing great promise as the basis
for Si light-emitting devices\cite{Reed04,Ossicini03}. The quantum
efficiency of light emission is significantly increased in Si QDs
compared with bulk Si because of the three-dimensional confinement
of carriers in small volumes ($< \sim$ 5 nm in diameter) that are
virtually free of defects. These QDs are usually embedded in a
SiO$_{2}$ matrix to take advantage of the high quality and
stability of the Si/SiO$_{2}$ interface. The dielectric nature of
SiO$_{2}$ requires that Si QDs are generally excited by hot
carriers when subjected to electrical pumping\cite{Franzo02}.

It has been shown that electrons are the dominant carriers in
silicon oxide containing Si QDs\cite{DiMaria83, DiMaria77}. The
incorporation of Si QDs leads to a conductivity larger than that
of SiO$_{2}$\cite{DiMaria83, DiMaria84}. DiMaria \emph{et
al}\cite{DiMaria83} examined all possible mechanisms for charge
transport in silicon oxide containing Si QDs, fabricated by
annealing Si-rich silicon oxide (SRSO) designated as SiO$_{x}$
(\emph{x} $<$ 2) with Si atomic concentrations of 34 - 39\% at
1000$\ ^\circ$C for 30 min. They concluded that the dominant
conduction mechanism appeared to be controlled by the tunneling of
electrons between the QDs. In another study, Maeda \emph{et
al}\cite{Maeda99} investigated the electrical properties of QDs
fabricated by the rapid thermal oxidation of an ultrathin
amorphous Si ($\alpha$-Si:H) film. They found that the charging
and consequent screening of the QDs led to \emph{N}-shaped
current(\emph{I})-voltage(\emph{V}) curves. These above examples
of work highlight the complex nature of electrical conduction in
silicon oxide containing Si QDs. Despite these and other
studies\cite{Torre03,Castagna03}, a complete picture has yet to be
proposed and fully described.

It is well known that electrons can travel through a thin
SiO$_{2}$ film by means of direct tunneling or Fowler-Nordheim
tunneling\cite{Depas95}. Direct tunneling indicates the presence
of a trapezoidal potential barrier whereas Fowler-Nordheim
tunneling takes place when electrons tunnel through a potential
barrier triangular in shape. Generally speaking, a large electric
field is needed to transform a rectangular potential barrier to a
triangular potential barrier. Therefore, Fowler-Nordheim tunneling
usually dominates in relatively high electric fields while direct
tunneling is the main conduction mechanism in low electric fields.
Although tunneling is regarded as being mainly responsible for
charge transport in SiO$_{2}$, the Frenkel-Poole effect has also
been observed in SiO$_{2}$\cite{Harrell99,Verwey72}.  The
Frenkel-Poole effect relates to the electric field enhanced
thermal emission of charge carriers from charged
traps\cite{Frenkel38,Harrell99}. Given the fact that Si QDs can be
charged\cite{Maeda99}, it is possible that they efficiently emit
charge carriers in silicon oxide with the help of electric fields.

In the present work \emph{I}-\emph{V} measurements were performed
at room temperature on an SRSO film with a Si atomic concentration
of 40\% after annealing at 750 - 1000$\ ^\circ$C. We focus on the
electrical conduction of silicon oxide in which Si QDs are already
electrically charged. It is found that either the Fowler-Nordheim
tunneling or the Frenkel-Poole effect can be used to explain the
electrical conduction in our experiment. We propose that at room
temperature Si QDs formed by Si-rich deposition and relatively low
temperature annealing can be regarded as traps that capture and
emit electrons by means of tunneling.

\section{Experimental details}
An SRSO film with a Si atomic concentration of 40\% was grown on a
$<$100$>$ Cz Si substrate (As-doped, \emph{n$^{+}$}-type, 0.0035
$\Omega$cm) at a temperature of 100$\ ^\circ$C by Electron
Cyclotron Resonance Plasma Enhanced Chemical Vapor Deposition
(ECR-PECVD). The deposition system has been described in detail
elsewhere\cite{Boudreau93}. The thickness and refractive index of
the film were determined with an ellipsometer to be 77 nm and
1.68, respectively. The film was cleaved into three samples
\emph{A}, \emph{B} and \emph{C}. Each sample was subsequently
annealed in a tube furnace with a flowing argon gas ambient for 3
h at 750, 900 and 1000$\ ^\circ$C, respectively. All the samples
were further annealed at 750$\ ^\circ$C for 1 h in the same
furnace with a flowing hydrogen gas ambient. Electrical backside
contacts were formed via 200-nm thick Al deposition followed by a
rapid thermal annealing at 550$\ ^\circ$C for 5 min in a nitrogen
ambient. Finally, 100-nm thick indium tin oxide (ITO) dots with a
diameter of 1.2 mm and a transmittance of $\sim$ 90\% in the
near-infrared region were deposited on the top of SRSO surface.

Photoluminescence (PL) from each sample was measured at room
temperature before the ITO deposition. The PL setup comprised a
Cd-He laser operating at a wavelength of 325 nm and an Ocean
Optics S2000 spectrometer that featured a high-sensitivity linear
CCD array . The effective power density of the laser beam on the
surface of the samples was $\sim$ 0.64 W/cm$^{2}$. Room
temperature \emph{I}-\emph{V} measurements were performed using an
Agilent 6624A DC power supply automatically controlled with a
Labview graphical user interface (GUI). The voltage was ramped
using a step of 0.5 V at a ramp rate of 1.9 V/s. The
\emph{I}-\emph{V} curve for each device was derived from the
measured voltages for an external series resistor of 100 $\Omega$.
The delay time for measurements after each voltage was applied was
5 s. This short delay led to hysteretic \emph{N}-shaped
\emph{I}-\emph{V} curves when all the devices were measured for
the first time, characteristic of the charging and screening of Si
QDs\cite{Maeda99}. A second measurement was carried out to obtain
the static \emph{I}-\emph{V} features immediately after a
hysteretic \emph{N}-shaped \emph{I}-\emph{V} curve was measured.
The absence of the hysteresis in the second measurement indicated
that Si QDs were already charged after the first
measurement\cite{Maeda99}.

\section{Results and discussion}
The existence of Si QDs in all the samples is evidenced by their
PL. Figure 1 shows the PL spectra for sample \emph{C} after
various annealing treatments. These spectra have been corrected
with respect to the system response of our PL setup. Si
nanocrystals (Si QDs) whose PL peaks at 827 nm are formed after
annealing at 1000$\ ^\circ$C in an argon ambient. After
hydrogenation at 750$\ ^\circ$C the PL intensity increases and the
PL peak redshifts to 842 nm. Both the redshift and the increase in
PL intensity are partially reversed following annealing at 550 $\
^\circ$C in a nitrogen ambient. These changes of PL intensity and
peak position are associated with the effect of impurities such as
hydrogen, which is discussed in detail elsewhere\cite{Pi06}. It is
important to note that after the formation of the nanocrystals
during annealing in an argon ambient their size changes only
marginally during annealing at 750$\ ^\circ$C in a hydrogen
ambient and at 550$\ ^\circ$C in a nitrogen ambient (this applies
to samples \emph{A} and \emph{B} also). X-ray diffraction
measurements suggest that Si nanocrystals with a characteristic PL
peak at 827 nm are 3.5 $\pm$ 0.7 nm in diameter.

Much weaker PL signals are observed from samples \emph{A} and
\emph{B} (not shown). This is caused by the amorphous nature of Si
QDs formed at temperatures below the crystallization temperature
$\sim$ 1000$\ ^\circ$C for nanometer-sized Si in silicon
oxide\cite{Pi06,Yi02,Molinari04,Iacona04}. Comparing the PL
spectra for samples \emph{A}, \emph{B} and \emph{C} we find that
the PL peak redshifts with the increase of annealing temperature
in an argon ambient. Assuming that quantum confinement is the
dominant mechanism for the PL from both amorphous and crystalline
Si QDs\cite{Molinari04}, we conclude that the mean size of the
dots in sample \emph{B} is larger than in sample \emph{A}, but
smaller than in sample \emph{C}.

No electroluminescence from these samples was detected at room
temperature with our Ocean Optics S2000 spectrometer. This does
not significantly affect the discussion on the electrical
conduction of our silicon oxide film containing Si QDs. The
absence of electroluminescence will be addressed subsequently in
conjunction with the results derived from the \emph{I}-\emph{V}
measurements. Figure 2 shows static current density \emph{J} as a
function of voltage \emph{V} for all the samples, in which Si QDs
are already charged. It is seen that \emph{J} decreases for a
given voltage with the increase of annealing temperature in an
argon ambient. Assuming a homogeneous distribution of QDs with a
negligible size dispersion, we relate the thickness \emph{s} of
oxide between the dots to the size \emph{d} of the dots, such that
\emph{s} = $(2.9687 \times 10^{7}/\sqrt[3]{C_{excess}}-1)d$, where
\emph{C$_{excess}$} is the excess Si concentration in
\emph{cm$^{-3}$}\cite{Zunger96}. For our film with a Si atomic
concentration of 40\% \emph{s} = 0.78\emph{d}. Therefore, \emph{J}
decreases with the increase of \emph{s}. This is consistent with a
tunneling effect, where the tunneling transmission probability of
electrons decreases with the increase of barrier width (s).

The breakdown voltage $\sim$ 31 V for sample \emph{C} is larger
than those for samples \emph{A} and \emph{B} $\sim$ 22 V, as a
result of their relative \emph{s} values. The relatively large
reverse bias of 31 V causes the formation of a depletion layer in
the near interface region of the Si substrate. The depletion layer
width can be up to 10 nm in our case\cite{Sze02}, which is large
enough to make the tunneling current density smaller under reverse
bias than under forward bias for sample \emph{C} (Fig. 2). We
further suggest that the corresponding depletion layer in sample
\emph{A} or \emph{B} below - 22 V is not wide enough to generate a
rectifying effect.

We assume the Fowler-Nordheim tunneling
mechanism\cite{Lenzlinger69} for the electrical conduction in all
the samples, in which the current density is given by
\begin{center}
\emph{J} =
\emph{fE$^{2}$q$^{3}$exp}[-\emph{4}(\emph{2m$^{*}$})\emph{$^{1/2}$$\phi^{3/2}$}/(\emph{3$\hbar$qE})]/(\emph{16$\pi^{2}\hbar\phi$}),\
\ \ \ (1)
\end{center}
where \emph{f}, \emph{E}, \emph{$\phi$}, \emph{m$^{*}$} and
\emph{q} are a correction factor, the electric field in the oxide
between Si QDs, the barrier height for electrons tunneling through
the oxide between Si QDs, effective electron mass and electron
charge, respectively. We plot \emph{$ln(J/E^2)$} against
1/\emph{E} under reverse bias in figure 3. In the calculation of
\emph{E} it is presumed that the voltage drops across the ITO, Si
QDs and the \emph{n$^{+}$}-type Si substrate, and the work
function difference between the ITO and \emph{n$^{+}$}-type Si
substrate, and the surface potential for the substrate-oxide
interface are all relatively small if not negligible. Thus,
\emph{E} = \emph{E$_{exp}$}(\emph{d} + \emph{s})/\emph{s}, where
\emph{E$_{exp}$} is the voltage drop across a device divided by
the film thickness of 77 nm. The solid lines are least square fits
of the form of equation (1). It is clear that both the critical
electric field \emph{E$_{FN}$} for the onset of Fowler-Nordheim
tunneling and the Fowler-Nordheim tunneling current density
decrease with the increase of annealing temperature in argon
during Si QDs formation. This is consistent with the dependence of
the Fowler-Nordheim tunneling on the barrier width
\emph{s}\cite{Depas95}. Electrical conduction in an electric field
below \emph{E$_{FN}$} is attributed to direct tunneling.

We calculate the effective barrier height $\phi_{eff}$ to be
$\sim$ 0.7 $\pm$ 0.1 eV by using \emph{E$_{FN}$} = 2.5 MV/cm and
\emph{s} = 2.7 $\pm$ 0.5 nm for sample \emph{C}. We note that
$\phi_{eff}$ is smaller than the barrier height $\sim$ 3 eV
obtained from photoconductivity and photoionization
measurements\cite{DiMaria83}. Figure 4 schematically shows the
energy levels for electrons in the present system. We believe that
the screening effect of charged Si QDs gives rise to localized
electric fields, transforming the original rectangular barriers
(Fig. 4 (a)) into trapeziform barriers (Fig. 4 (b)). For the
occurrence of the Fowler-Nordheim tunneling (Fig. 4 (c)) electrons
need the help of external fields to overcome the smaller bases of
these trapezoids, which are 0.7 eV $\pm$ 0.1 eV, similar to the
value of 0.6 eV obtained by DiMaria \emph{et al}\cite{DiMaria83}.
It is shown in figure 2 that \emph{J} very weakly depends on the
electric field polarity if the effect of the depletion layer is
not considered. This results from the fact that the localized
electric fields caused by charged Si QDs change direction with
respect to the external electric fields.

Assuming that $\phi_{eff}$ is nearly the same for all the samples,
we calculate the values of \emph{s} to be 1.6 $\pm$ 0.2 and 1.9
$\pm$ 0.3 nm for samples \emph{A} and \emph{B}, respectively.
Using these values of \emph{s} we can estimate the maximum kinetic
energy (\emph{K.E.}) of an electron as it moves between
neighboring QDs. In the simple one-dimensional case \emph{K.E.} =
\emph{Eqs}. For samples \emph{A} and \emph{B} the maximum
\emph{K.E.} (determined by the film electrical breakdown) is less
than 1.5 eV, which is not large enough to excite an electron-hole
pair according to the PL data. For sample \emph{C}, however, where
a breakdown field of 9 MV/cm is achievable, the \emph{K.E.} could
reach 2.4 eV, presumably enough to cause electroluminescence. The
absence of electroluminescence from sample \emph{C} then appears
related to the small current of electrons able to tunnel between
neighboring QDs (Fig. 2).

We have also considered the Frenkel-Poole effect\cite{Frenkel38,
Harrell99} as a model for charge transport in all the samples, in
which the current density is given by
\begin{center}
\emph{J} =
\emph{CEexp}\{-\emph{q$\psi$}/(\emph{2k$_{B}$T})+[\emph{q$^{3}$E}/(\emph{$\pi\varepsilon$})]\emph{$^{1/2}$}/(\emph{2k$_{B}$T})\},\
\ \ \ (2)
\end{center}
where \emph{C}, \emph{$\psi$}, \emph{k$_{B}$},
\emph{$\varepsilon$} and \emph{T} are a system specific constant,
the barrier height for trapped electrons to escape from Si QDs,
the Boltzmann constant, the dielectric constant of the film and
temperature, respectively. In Figure 5 we plot \emph{$ln(J/E)$}
against \emph{E$^{1/2}$} under reverse bias. The solid lines are
least square fits based on the Frenkel-Poole effect. The critical
electric field \emph{E$_{FP}$} for the beginning of the
Frenkel-Poole effect approximates \emph{E$_{FN}$} for each sample.
Using \emph{$\psi_{eff}$} = \emph{qs}\emph{E$_{FP}$}, where
\emph{$\psi_{eff}$} is the effective barrier height for trapped
electrons to escape from Si QDs we estimate from sample \emph{C}
that \emph{$\psi_{eff}$} is 0.7 $\pm$ 0.1 eV. The same values of
\emph{$\psi_{eff}$} and \emph{$\phi_{eff}$} lead us to believe
that at room temperature Si QDs can be regarded as traps that
capture and emit electrons by means of tunneling. The refractive
index \emph{n} (\emph{n} = $\varepsilon^{1/2}$) has been estimated
from the slope of each Frenkel-Poole effect fitting. The values of
\emph{n} are $1.74 \pm 0.07$, $1.85 \pm 0.26$ and $3.55 \pm 1.85$
for samples \emph{A}, \emph{B} and \emph{C}, respectively. The Si
QDs induced increase of refractive index of silicon oxide has been
observed previously\cite{Pavesi00}; however the value of 3.55 for
sample \emph{C} is unrealistic and difficult to estimate with
greater accuracy due to the small measured current density.

\section{Conclusion}
In summary, current-voltage measurements have been made at room
temperature on identically deposited SRSO films annealed at 750 -
1000$\ ^\circ$C. The thickness of oxide between the Si QDs
embedded in the film increases with the increase of annealing
temperature. This leads to a decrease in current density as the
annealing temperature is increased. Assuming the Fowler-Nordheim
tunneling mechanism in large electric fields, we obtain an
effective barrier height $\phi_{eff}$ of $\sim$ 0.7 $\pm$ 0.1 eV
for an electron tunneling through an oxide layer between Si QDs.
The Frenkel-Poole type behavior can also be used to explain the
electrical conduction of the film in large electric fields. It is
suggested that at room temperature Si QDs can be regarded as traps
that capture and emit electrons by means of tunneling.

\textbf{Acknowledgements}\\
The authors thank Dr. A. Kitai and Mr P Jonasson for assistance in
device fabrication. X-ray diffraction measurements are provided
courtesy of Dr D. Comedi. This work is supported by the Natural
Sciences and Engineering Research Council of Canada, Ontario
Centers of Excellence Inc., and the Ontario Research and
Development Challenge Fund under the Ontario Photonics Consortium.

\newpage
FIG. 1:  The PL spectra for sample \emph{C}, which is (a)
initially annealed at 1000$\ ^\circ$C for 3 h in an argon ambient
and (b) further at 750$\ ^\circ$C for 1 h in a hydrogen ambient
and (c) finally at 550$\ ^\circ$C for 5 min in a nitrogen
ambient.\vspace{2cm}

FIG. 2:  The current density as a function of voltage for samples
\emph{A}, \emph{B} and \emph{C}, which are initially annealed in
an argon ambient for 3 h at 750, 900 and 1000$\ ^\circ$C,
respectively. The current density for sample \emph{C} is shown 80
times larger than measured.\vspace{2cm}

FIG. 3:  The Fowler-Nordheim plots for samples \emph{A}, \emph{B}
and \emph{C}, which are initially annealed in an argon ambient for
3 h at 750, 900 and 1000$\ ^\circ$C, respectively. The solid lines
are least square fits to the Fowler-Nordheim tunneling. \emph{E}
and \emph{J} are in V/cm and A/cm$^{2}$, respectively.
\vspace{2cm}

FIG. 4:  The schematic of energy levels for electrons in silicon
oxide containing Si QDs. (a) The original barrier is rectangular
with a barrier height of $\sim$ 3 eV. (b) The localized electric
field induced by the screening effect of charged Si QDs transforms
the rectangular barrier to a trapezoidal barrier. (c) The
Fowler-Nordheim tunneling occurs when electrons gain energy to
overcome the smaller base ($\phi_{eff}$) of the trapezoidal
barrier with the help of the external field \emph{E}.\vspace{2cm}

FIG. 5:  The Frenkel-Poole plots for samples \emph{A}, \emph{B}
and \emph{C}, which are initially annealed in an argon ambient for
3 h at 750, 900 and 1000$\ ^\circ$C, respectively. The solid lines
are least square fits to the Frenkel-Poole effect. \emph{E} and
\emph{J} are in V/cm and A/cm$^{2}$, respectively.

\newpage
\begin{figure}[h]\centering
\scalebox{0.4}[0.4]{\includegraphics*[47,722][762,215]{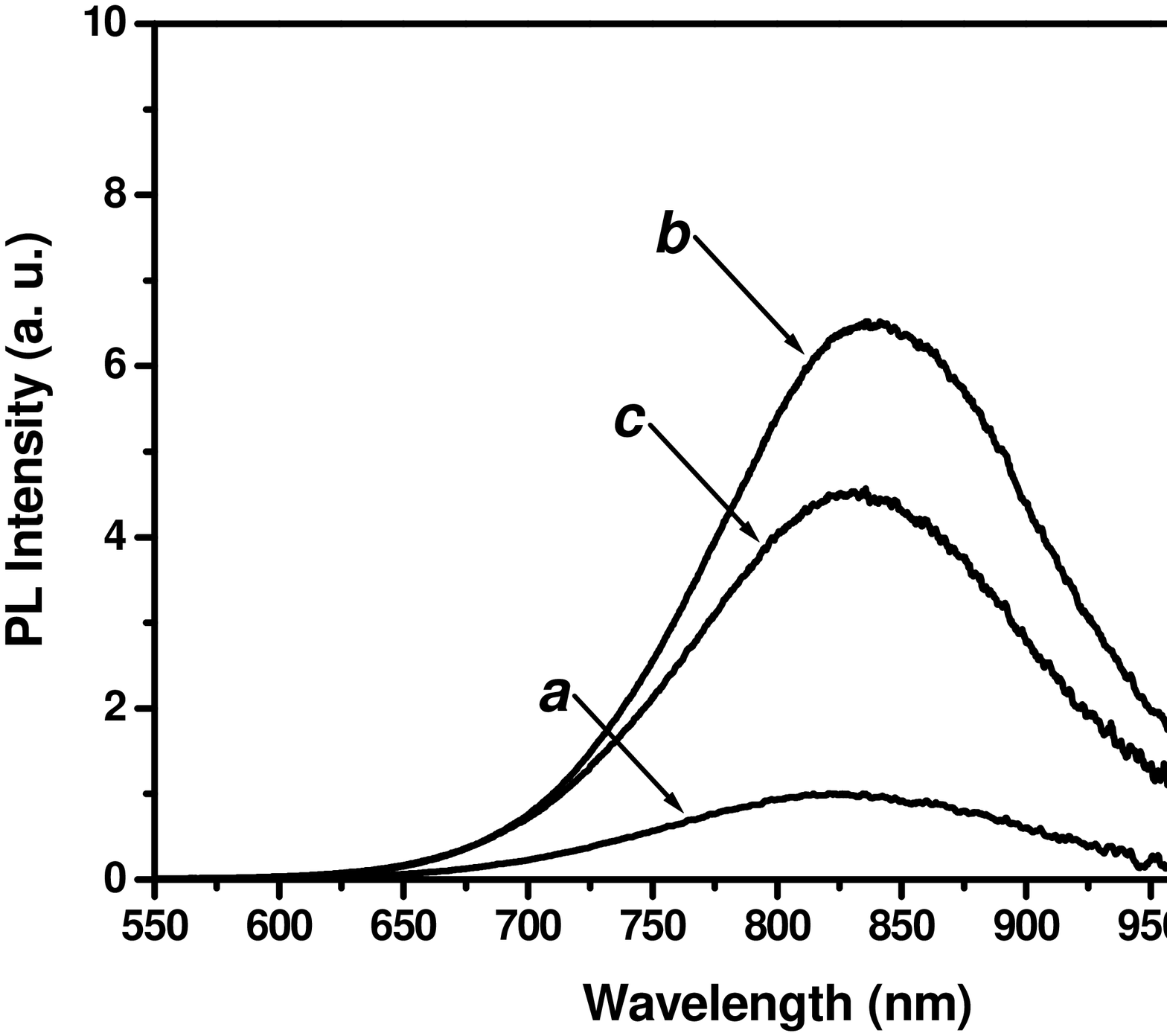}}
\vspace{15cm} \caption{\label{Fig1}Pi \emph{et} \emph{al}'s}
\end{figure}

\newpage
\begin{figure}[h]\centering
\scalebox{0.4}[0.4]{\includegraphics*[44,731][716,229]{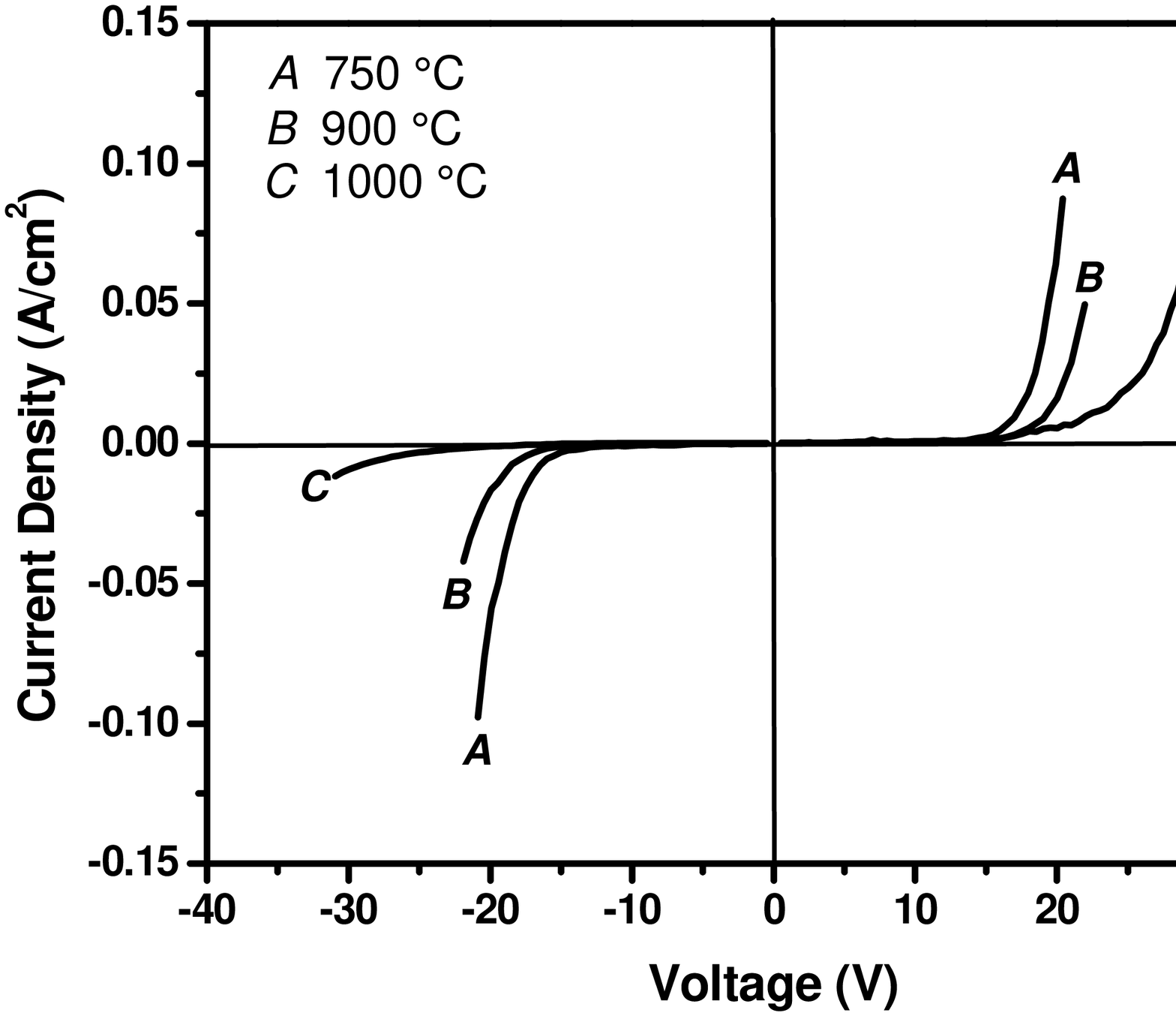}}
\vspace{15cm} \caption{\label{Fig2}Pi \emph{et} \emph{al}'s}
\end{figure}

\newpage
\begin{figure}[t,b,h]\centering
\scalebox {0.4}[0.4]{\includegraphics*[65,724][737,224]{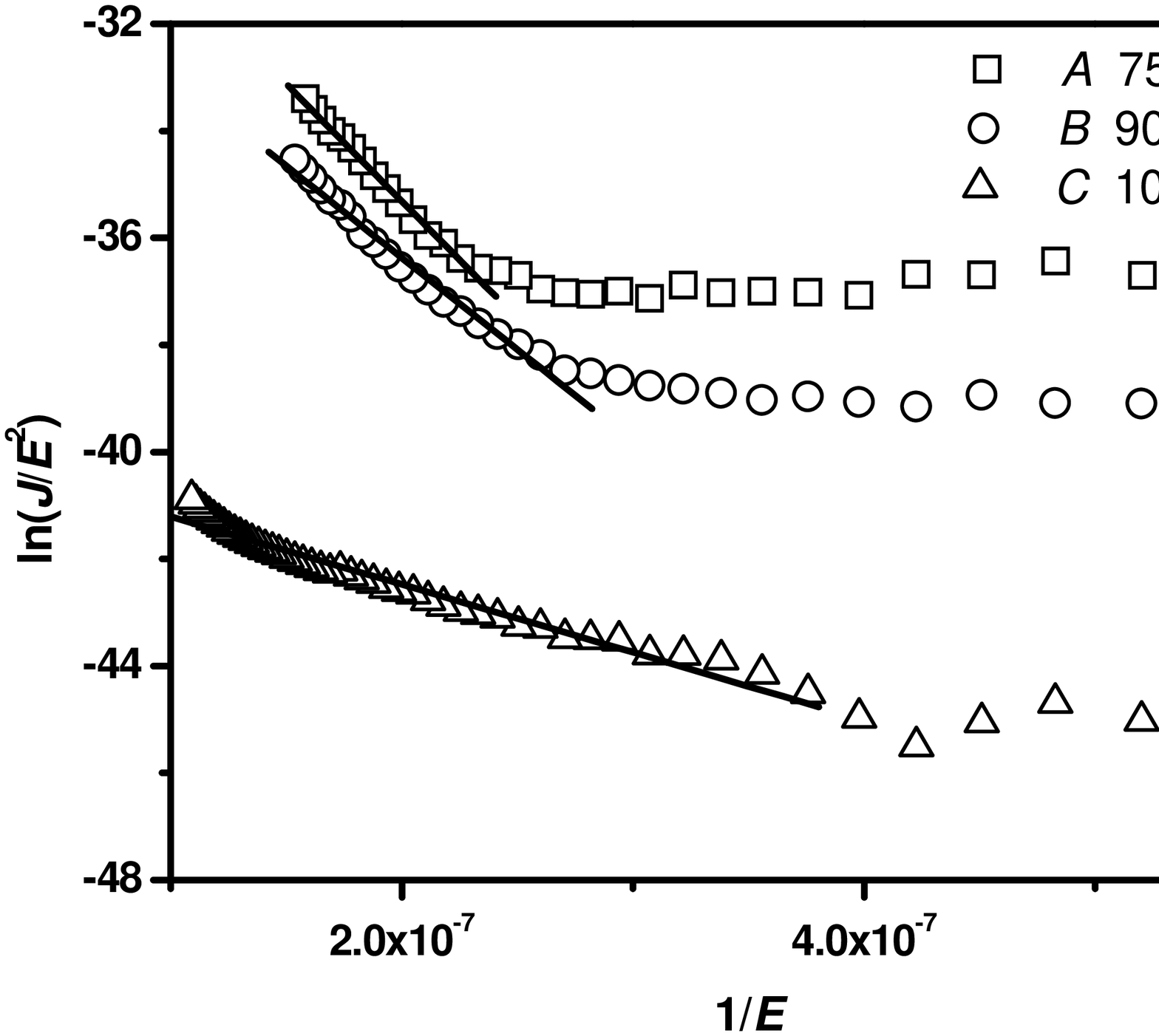}}
\vspace{15cm} \caption{\label{Fig3}Pi \emph{et} \emph{al}'s}
\end{figure}

\newpage
\begin{figure}[h]\centering
\scalebox{.72}[.72]{\includegraphics*[95,548][448,273]{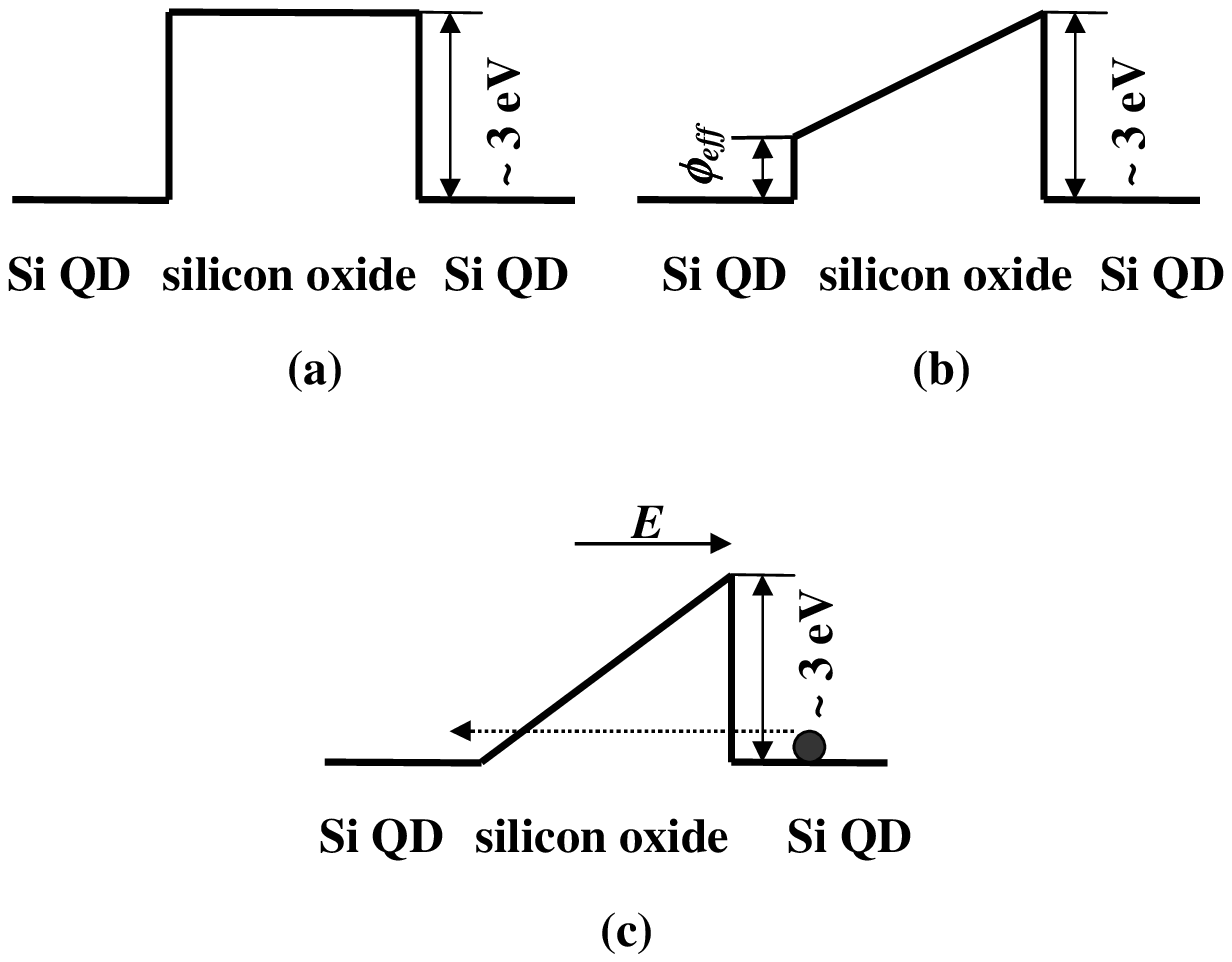}}
\vspace{15cm} \caption{\label{Fig4}Pi \emph{et} \emph{al}'s}
\end{figure}

\newpage
\begin{figure}[t,b,h]\centering
\scalebox {0.4}[0.4]{\includegraphics*[67,727][723,223]{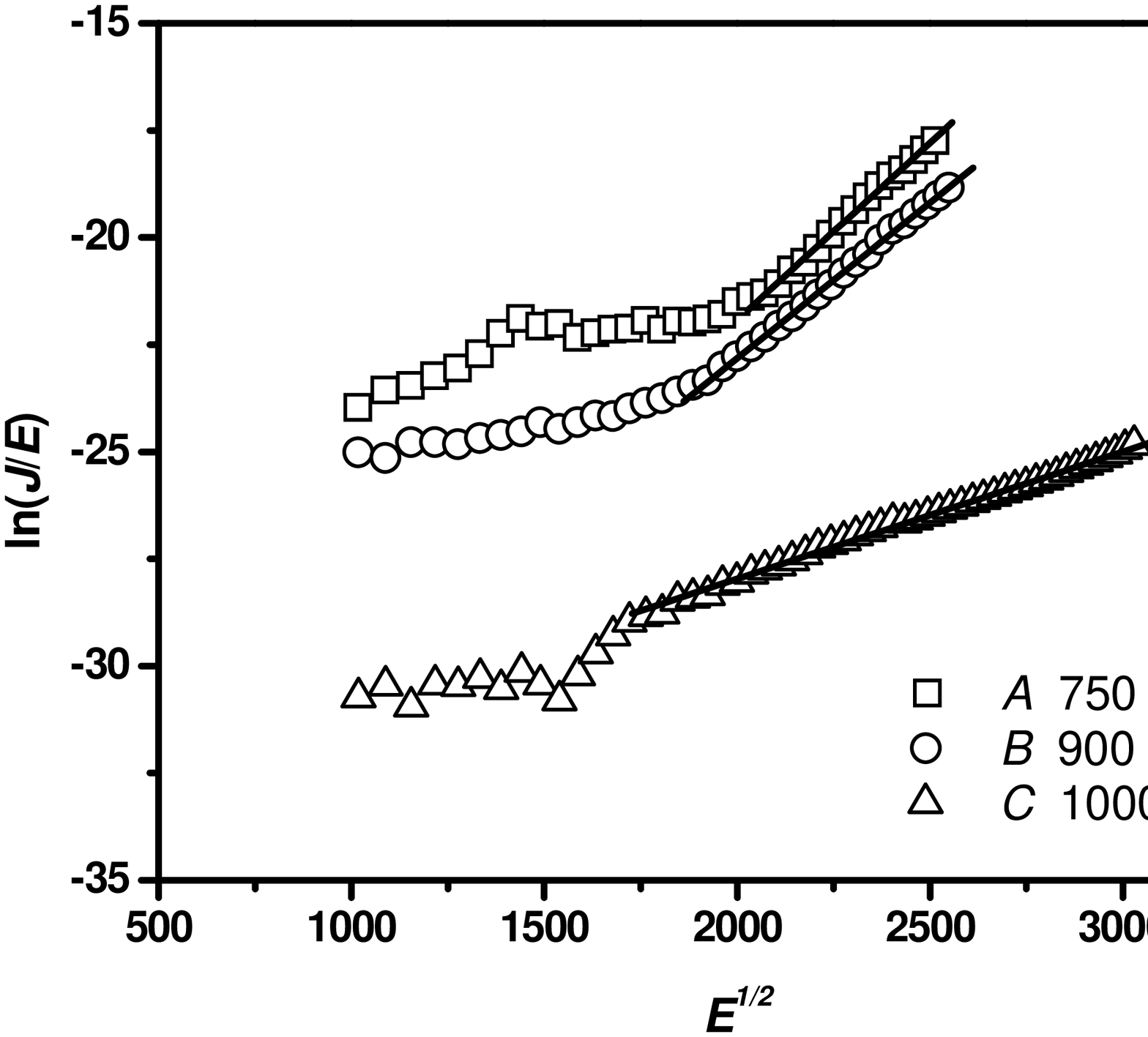}}
\vspace{15cm} \caption{\label{Fig5}Pi \emph{et} \emph{al}'s}
\end{figure}

\end{document}